\documentstyle[aps,prb,preprint]{revtex}
\tightenlines
\begin{document}

\include{psfig}

\draft \title{\bf Frequency-dependent fluctuational conductivity above 
T$_{c}$ in anisotropic superconductors: effects of a short wavelength 
cutoff}

\author{E.Silva$^{(1)}$}

\address{ $^{(1)}$Dipartimento di Fisica ``E.Amaldi'' and Unit\`{a} 
INFM, Universit\`{a} di Roma Tre, Via della Vasca Navale 84, 00146 
Roma, Italy}

\date{European Physical Journal B 27, 497 (2002)} \maketitle

\begin{abstract}
We discuss the excess conductivity at nonzero frequencies in a 
superconductor above $T_c$ within the gaussian approximation. We focus the attention on the 
temperature range not too close to $T_{c}$: within a time-dependent 
Ginzburg-Landau formulation, we phenomenologically introduce a short 
wavelength cutoff (of the order of the inverse coherence length) in 
the fluctuational spectrum to suppress high momentum modes.  We treat 
the general cases of thin wires, anisotropic thin films and 
anisotropic bulk samples.  We obtain in all cases explicit expressions 
for the finite frequency fluctuational conductivity. The dc 
case directly follows. Close to $T_{c}$ 
the cutoff has no effect, and the known results for Gaussian 
fluctuations are recovered.  Above $T_{c}$, and already for 
$\epsilon=ln(T/T_{c})>10^{-2}$, we find strong suppression of the 
paraconductivity as compared to the gaussian prediction, in particular 
in the real part of the paraconductivity.  At high $\epsilon$ the 
cutoff effects are dominant. We discuss our results in comparison 
with data on high-$T_{c}$ superconductors.
\end{abstract}
\pacs{74.40.+k\\ 74.76.-w\\78.70.Gq}

\section{Introduction}
\label{intro}

After many years of studies of the fluctuational effects on various 
properties of superconductors, the discovery of high-$T_{c}$ 
superconductors (HTC) has given new impulse to the study of 
fluctuational effects above the critical temperature $T_{c}$.  In 
fact, high critical temperatures, short coherence lengths and strong 
anisotropy greatly enhance the temperature range of observation of 
fluctuational effects.  This fact has led to the search for extensions 
of theoretical models, such as Ginzburg-Landau (GL) theory, to 
temperature ranges even well above $T_{c}$.\cite{mishonov}\\
In general, fluctuational effects have been widely investigated in the 
past, both in conventional, low-$T_{c}$ (LTC) 
superconductors\cite{skoc} and in high-$T_{c}$ cuprates (HTC).  Among 
the various effects affected by thermal fluctuations, the dc excess 
conductivity above $T_{c}$ has been experimentally studied most 
extensively.\cite{freitas,ausloos,hopf,balest92,labdi,calzona,pavuna} 
Theoretical investigations dates from the 
sixties,\cite{skoc,asla,schmidt,maki,klemm} and development of this 
topic proceeded until the discovery of HTCs.  With some noticeable 
exception,\cite{mishonov,freitas,hopf,reggiani,gauzzi,silvaPhC,silva} 
the attention was always focussed in the temperature region close to 
the critical temperature.  In that region, it is customary to discuss 
the data in terms of the Aslamazov-Larkin\cite{asla} (AL) and 
Maki-Thompson\cite{maki,thom} contributions.  Alternatively, the data are 
often described in terms of gaussian fluctuations, which by the 
simpler formalism of Ginzburg-Landau theory yield the same results as 
the microscopic AL approach\cite{mishonov,maki} in clean 
superconductors.\\
While such treatments are in principle valid only close to 
$T_{c}$, in HTCs the range of observation 
of fluctuational effects is much wider. Thus, analysis of the data 
only in the strict proximity of $T_{c}$ would miss a substantial 
amount of information, eventually leading to incomplete 
conclusions.  Tools for investigating a wider temperature range are 
then not only interesting by themselves, but necessary for the study 
of HTC compounds.  \\
Hystorically, it was recognized very early \cite{johnson2,johnson} 
that the experimental dc excess conductivity dropped much faster than 
predicted by, e.g., AL theory,\cite{asla} when temperature was raised 
above $T_{c}$.  Early works\cite{johnson,werthamer} attributed this 
behavior to the unphysical contribution of high momentum modes to the 
fluctuation conductivity: close to $T_{c}$ the uniform mode ($q=0$) is 
the most diverging one, so that the contributions of the other modes 
are automatically negligible, but at even slightly higher temperatures 
this is no longer true, and one has to manage somehow the other 
contributions.  An essentially similar framework, both in the 
experimental results and theoretical arguments, applied to the 
fluctuation diamagnetism above $T_{c}$.\cite{patton,gollub2,gollub} 
In fact, the relevance of short or long wavelength modes is a rather 
general issue of the physics of phase transitions: it is here 
interesting to note that arguments similar to those developed in this 
paper were raised, e.g., in the treatment of the electrical resistivity 
in magnetic systems.\cite{magnetic}\\
In most theoretical calculations a phenomenological approach was used: 
the dc excess conductivity was calculated within the time-dependent 
Ginzburg-Landau (TDGL) theory.  Bearing in mind that within this 
approach slow variations of the order parameter (on the scale of 
$\xi_{0}$, the coherence length) are 
required,\cite{mishonov,werthamer} so that short-wavelength 
fluctuations are unphysical, a high-$q$ cutoff was 
imposed.\cite{mishonov,hopf,pavuna,gauzzi,silvaPhC,silva,johnson,nerisatt,nerimos,coffey} 
A numerical approach was almost everywhere used, apart from a few 
analytical results in some special 
cases.\cite{pavuna,gauzzi,silvaPhC,silva,coffey}\\
A second approach, grounded on microscopic calculations, was employed 
in early works to obtain the flluctuation 
diamagnetism.\cite{makitaka,kurk,gerhardts} The calculation of 
fluctuational conductivity was carried out by A.Varlamov and 
co-workers in several papers,\cite{varlareview} within the 
Lawrence-Doniach (LD) model for layered superconductors (that in the 
high temperature range reduces to a collection of two-dimensional 
superconducting sheets).  In this case the basis was the calculation 
of the full energy and momentum dependence of the fluctuation 
propagator and of the Green functions.\cite{reggiani} The dc 
conductivity was then calculated also for temperatures well above 
$T_{c}$.\\
Both these approaches gave as a result a strongly reduced dc excess 
conductivity well above $T_{c}$.  A detailed comparison between the 
microscopic and the cutoffed, GL model for clean 2D superconductors 
has been given in Ref.\onlinecite{silva}, where it was shown that, with an 
appropriate choice of the cutoff, the two formulations gave the same 
temperature dependence of $\Delta\sigma$.\\
Comparisons with experiments were made on various materials: the 
phenomenological approach was successfully compared to data for the dc 
excess conductivity in amorphous alloys\cite{johnson} and 
YBa$_2$Cu$_3$O$_{7-\delta}$ (YBCO) films\cite{hopf,pavuna} and 
crystals\cite{carballeira} within a 
three-dimensional model, and in Bi$_2$Sr$_2$CaCu$_2$O$_{8+x}$ (BSCCO) 
crystals\cite{silvaPhC,silva} and tapes\cite{wang} with a 
two-dimensional expression. The 2D limit of the microscopic 
calculation\cite{reggiani} was shown to be in agreement with the data 
in BSCCO films.\cite{balest92}  In all cases the agreement extended 
at least up to $T\simeq 1.25 T_{c}$. It is worth mentioning that a doping 
dependence of the cutoff number was found in BSCCO 
crystals,\cite{silva} thus opening interesting issues about possible 
links between the phenomenological cutoff approach and microscopic 
properties.\\
In addition to the studies of the dc conductivity, a few papers 
appeared which dealt with the finite-frequency excess conductivity, 
mainly in the microwave 
range,\cite{nerisatt,nerimos,oldmicro,anlage,booth,neri,waldram} 
together with the corresponding theoretical 
treatments.\cite{skoc,schmidt,klemm,dorsey} In general, early 
experimental works\cite{oldmicro} were found to agree reasonably well 
with pure gaussian models,\cite{schmidt} while the data in HTC showed 
a possible evidence for critical behavior close to 
$T_{c}$.\cite{booth,neri} Differently from the dc case, only a very 
few preliminary works\cite{nerisatt,nerimos} explored the high 
temperature regime.  In these cases, extra conductivity lower than the 
gaussian expectation was experimentally found, but no simple 
theoretical expressions were given.  It is interesting to note that 
none of the reported microwave studies has shown evidence for the 
Maki-Thompson term, neither in low-$T_{c}$ nor in high-$T_{c}$ 
superconductors, as opposed to some dc experiments.\cite{skoc,ausMT} 
In addition, recent microscopic theoretical results have shown that in 
the ultraclean limit the fluctuationl conductivity is entirely given 
by the AL term, since the other contributions cancel 
out.\cite{livanov} We find in these facts strong motivation for the 
present study.\\
It should be mentioned that the cutoff approach is still a debated 
topic.\cite{mishonov,silva,carballeira} However, it is a matter of 
fact that, to the lowest level, it gives a very good description of 
the data, while on fundamental grounds it is believed\cite{mishonov} to be a 
fundamental aspect of the GL theory.\\
In this paper we present an explicit calculation of the 
finite-frequency excess conductivity based upon the TDGL theory.  We 
phenomenologically introduce the short wavelength cutoff in the 
fluctuational spectrum.  Differently from previous results, we obtain 
closed forms for the dynamic excess conductivity.  Formulae are 
presented for thin wires (one-dimensional systems, 1D), thin films 
(2D) and bulk (3D) superconductors.  We explicitly include the 
in-plane anisotropy for the 2D case, and the three-axis anisotropy for 
the 3D case.  Our results extend previous calculations in several 
ways: {\em i)} to different dimensionalities, {\em ii)} including the 
frequency dependence, {\em iii)} giving explicit expressions for the 
cutoffed excess conductivity.\\
We discuss some of the features that can be expected in the 
experimental data when entering the short-wavelength regime.  We 
expect that our calculations might be useful for the description 
of the frequency-dependent excess conductivity not too close to 
$T_{c}$.

\section{Model}
\label{model}
We use TDGL theory in order to investigate the main effects of a 
short-wavelength cutoff on the finite frequency conductivity.  We work 
in absence of a dc magnetic field.  We calculate the current response 
to a plane-wave electric field ${\bf E}e^{-i\omega t}$.  Throughout 
the paper we use Sist\`{e}me International units for the ease of 
comparison of the results with the experiments.  The GL functional is 
written in the general case as:

\begin{equation}
F=\int d^{D}r\left[ \sum\limits_{j=1,..D}\frac{\hbar ^2}{2m_j}\left| 
\left( \frac \partial {\partial r_j}-\frac{ie^{\star}}\hbar A_j\right) 
\psi \left( {\bf r}\right) \right| ^2+\alpha \,\left| \psi \left( {\bf 
r}\right) \right| ^2+\frac 12\beta \,\left| \psi \left( {\bf r}\right) 
\right| ^4\right]
\label{th1}
\end{equation}

where the symbols have the following usual definitions:\\
The vector potential ${\bf A}=-\frac{i}{\omega} {\bf E}$;\\
$\alpha = a\epsilon$, where $\epsilon = \ln(T/T_c)$ is the reduced 
temperature; \cite{gorkov}\\
$e^{\star}=2e$ is twice the electronic charge;\\
$m_{j}$ are the masses of the pair along the main crystallographic 
directions.  Since we deal with the general case of anisotropic 
superconductors, all of them are in principle different.\\
The GL coherence lengths along the different axes are then defined as 
usual as $\xi_{j}=\hbar/\sqrt{2m_{j}\alpha}$\\
We are mainly interested in the behavior not too close to $T_{c}$, so 
that in the following we neglect the term $\sim\psi^{4}$ altogether 
(inclusion of this terms can be relevant very close to 
$T_{c}$\cite{neri,dorsey,mikeska}).  Our aim is to calculate the 
longitudinal dynamical conductivity along the axis of application of 
the ac field, chosen to be the $x$ axis.  In the following we will 
consider three distinct ``dimensionalities":\\
{\em i)} 1D: thin superconducting wire, which we take to lay along 
the $x$ axis.  The section of the wire is assumed to be a square of 
side $L$, with $L<\xi_{y},\xi_{z}$.  Only $m_{x}$, $\xi_{x}$ are 
relevant.\\
{\em ii)} 2D: thin superconducting film, which is taken to lay in the 
$x,y$ plane.  The thickness $d<\xi_{z}$.  Both $m_{x}$ ($\xi_{x}$) and 
$m_{y}$ ($\xi_{y}$) are relevant.\\
{\em iii)} 3D: bulk anisotropic superconductor.  The axes are taken 
along the crystallographic axes.  All three masses and coherence 
lengths are relevant.\\
We do not rewrite all the preliminary calculations, that can be 
found elsewhere.\cite{neri,dorsey} Instead, we sketch the path that 
has been followed.  The time evolution of the order parameter, needed 
in order to calculate the dynamical conductivity, is determined by the 
TDGL equation describing relaxational dynamic for $\psi$, which is 
taken to include a white noise term. At this stage, the 
characteristic relaxation rate $\Gamma_{0}$ is introduced. \cite{gamma0}\\
The response to the field ${\bf A}(t)$ is determined by the current 
operator averaged with respect to the noise (represented below by the 
brackets), and it can be expressed as a function of the correlation 
function of the order parameter $C\left( {\bf r},t;{\bf r}^{\prime 
},t^{\prime }\right) =\left\langle \psi \left( {\bf r},t\right) \;\psi 
^{\star}\left( {\bf r}^{\prime },t^{\prime }\right) \right\rangle $ at 
equal times:

\begin{equation}
\left\langle J_x\left( t\right) \right\rangle =- \frac{\hbar 
e^{\star}}{m_{x}}\int \frac{d^D q}{\left( 2\pi \right) ^D}q_xC\left[ 
{\bf k}={\bf q}- \frac {e^{\star}} {\hbar} {\bf A}\left( t\right) 
;t,t\right] \
\label{corrente}
\end{equation}

where the momentum dependence has been shifted from ${\bf k}$ to the 
new vector ${\bf q}={\bf k}+\left( e^{\star}/\hbar \right) {\bf 
A}\left( t\right) $.\\
The correlator has been calculated by, e.g., A.T.Dorsey\cite{dorsey} 
in the gaussian approximation.  We limit ourselves to the treatment of 
the linear response (we note however that the present approach has 
been used to calculate the nonlinear excess conductivity close to 
$T_{c}$, Ref.\onlinecite{dorsey}).  Within the linear response, all the quadratic 
terms in the vector potential can be neglected, and the equal time 
correlator reads:\cite{notamisho}

\begin{eqnarray}
C\left( {\bf q};t,t\right) =2k_BT\Gamma _0\int\limits_0^{+\infty }ds 
exp \left\{\left[ {-2\Gamma _0\alpha s-\Gamma _0\hbar ^2s\left( 
\frac{q_y^2}{m_{y}} +\frac{q_z^2}{m_z}\right)} \right]+ \right.
\label{correlatore}
\\
\left.  -\frac{\Gamma _0\hbar ^2s}{m_{x}} 
\left[q_x^2-\frac{2e^{\star}q_x}{\hbar s}\int\limits_0^sdu\left[ 
A_x\left( t-u\right) -A_x\left( t\right) \right] \right] \right\} 
\nonumber
\end{eqnarray}

The usual calculation is based on the integration of Eq.\ref{corrente} 
for all momenta.  Then, inserting Eq.\ref{correlatore} in 
Eq.\ref{corrente} and performing the integrals for the appropriate 
dimensionalities one gets the well known gaussian 
results.\cite{schmidt,dorsey}\\
As discussed in the Introduction, we wish to phenomenologically extend 
the applicability range of the calculation to temperatures not too 
close to $T_{c}$, so that we explicitly discard the short-wavelength 
fluctuations by inserting a high-$q$ cutoff in the integration of 
Eq.\ref{corrente}.  There are several possible choices for this task.  
In the calculation of the dc conductivity with the aim of the Kubo 
formula, Gauzzi\cite{gauzzi} has chosen a different cutoff for each 
$q$-component.  Hopfengartner et al.\cite{hopf} have chosen, for the 
special case of YBCO taken as a uniaxial superconductor, the same 
cutoff for $q_{x}$ and $q_{y}$, and a different one on $q_{z}$.  
Johnson et al.\cite{johnson} used a cutoff on the modulus of $q$ in 
order to describe results on 3D, isotropic superconductors.\\
It is worth stressing that in the comparison with the experiments, 
each cutoff becomes a new fitting parameter.  It is in general 
important to reduce the number of such parameters, and particularly in 
a calculation like this, where phenomenological modifications are 
made.  To meet this requirement, it is useful to note that on physical 
grounds the cutoff is measured in units of the inverse 
temperature-independent correlation length,\cite{werthamer,notaxi} 
$\xi_{0}^{-1}$.  In homogeneous superconductors, even if anisotropic, 
there are no evident reasons for which a different number of 
$\xi_{0}^{-1}$ should be used as a cutoff in different directions.  We 
then introduce a single cutoff such that 
$\sqrt{\sum\limits_{j=1,..D}[q_j\xi_j(0)]^{2}}<\Lambda$ in 
Eq.\ref{corrente}.  We mention that this choice might be questionable 
when explicitly dealing with layered superconductors, where there is 
indeed a physical difference for in-plane and out-of-plane properties.  
This would require the use of the Lawrence-Doniach\cite{ld} or similar 
model, which is not the purpose of this work.  A study limited to the 
dc conductivity can be found in Refs.\cite{hopf,coffey}, while for a 
more general formulation of the extended (cutoffed) GL theory applied 
to layered superconductors we address the reader to a recent 
review.\cite{mishonov}\\
To perform the calculations it is useful to manipulate the expression 
for the current using dimensionless variables: \\
$k_{j}=q_{j}\xi_{j}$, with $j=x,y,z$, and\\
$w=\frac{\omega}{2\Gamma_{0}\alpha}=\omega\tau=\frac{\omega\tau_{0}}{\epsilon}$\\
where in the latter relation the second equality defines the 
temperature-dependent GL characteristic time $\tau$, and the 
third equality defines $\tau_{0}$.\cite{gamma0} In these units the 
cutoff is expressed as $K=\Lambda/\sqrt{\epsilon}$.  The full 
expression for the current becomes:

\begin{equation}
\left\langle J_x\left( t\right) \right\rangle 
=-\frac{2e^{*}k_{B}T}{\xi_{x}\xi_{y}\hbar}\int\limits_0^{+\infty }du 
e^{-u}\int\frac{d^{D}k}{(2\pi)^{D}}k_{x} exp 
\left\{\-uk^{2}+B_{x}\frac{E k_{x}}{w^{2}}\left[ {e^{iuw}-1-iuw} 
\right]\right\}
\label{Jnew}
\end{equation}

Where $B_{x}=e^{*}(2\alpha^{3}m_{x})^{-\frac{1}{2}}/\Gamma_{0}$.  This 
is the starting point for all the results presented in the following 
section.  It is intended that in 2D we make use of 
$\int\frac{d^{D}q}{(2\pi)^{D}}\rightarrow\frac{1}{d}\int\frac{d^{2}q}{(2\pi)^{2}}$, 
and in 1D 
$\int\frac{d^{D}q}{(2\pi)^{D}}\rightarrow\frac{1}{L^{2}}\int\frac{dq}{2\pi}$, 
that is we do not allow (consistently with the rest of the present 
work) high-$q$ modes along the film thickness or wire section.\\
We do not explicitly report the long, but trivial procedures, for each 
specific case.  In all cases we get, in the limit of small fields, the 
$x$-axis response written as:

\begin{equation}
\langle J_x\rangle =\left[ \sigma^\prime ( \omega ) 
+i\sigma^{\prime\prime} ( \omega)\right] \;E_{x}\;e^{-i\omega 
t}=\sigma(\omega)E_{x}\;e^{-i\omega t}
\label{conducibilita}
\end{equation}

which defines the longitudinal complex conductivity 
$\sigma=\sigma^{\prime}+i\sigma^{\prime\prime}$.  We recall that with 
the inclusion of all the momentum contributions, in 2D and 3D it was 
found\cite{schmidt,dorsey} that the gaussian conductivity was a 
scaling function of the frequency:

\begin{equation}
\sigma_g(\omega )=\sigma_g^{\prime }(\omega )+i\sigma_g^{\prime \prime 
}(\omega )=\sigma_{g}^{(dc)}\left[ S_{+}\left( \omega \tau
\right) +i\,S_{-}\left( \omega \tau \right) \right]
\label{sigmagau}
\end{equation}

where $S_+ (x)$ and $S_- (x)$ are the scaling functions as can be 
found in Ref.\onlinecite{dorsey} (with the subscript $ g $ we indicate 
the results obtained in the gaussian approximation without a momentum 
cutoff), and they have the property that $S_{+} (x\to 0)=1$ and $S_{-} 
(x\to 0)=0$.  The scaling expression has found some experimental 
confirmation in swept-frequency measurements in YBCO,\cite{booth} but 
the temperature dependence of $\tau$ was found to markedly depart from 
the GL prediction, $\tau\sim\epsilon^{-1}$.

\section{results}
\label{results}
In this Section we present the resulting expressions for the 
finite-frequency excess conductivity for bulk superconductors, thin 
films and thin wires.  In all cases, we find closed forms for the 
complex conductivity, which for compactness we leave in the form of 
functions of complex variable.  Calculation of Eq.\ref{Jnew} leads to 
the following expression for the anisotropic 3D complex conductivity:
\begin{eqnarray}
	\sigma_{3D}=\frac{e^{2}}{32\hbar \xi_z(0)\epsilon^{1/2}} 
	\;\frac{\xi_x(0)}{\xi_y(0)}\;\frac{16}{3\pi w^{2}} \left\{atn\; K 
	-\left(1-iw\right)^{3/2}\;atn\left(\frac{K}{\sqrt{1-iw}}\right)+ 
	\right.  \nonumber \\
	\left.  +iw\left[ \frac{K}{2\left(1+K^{2}\right)}-\frac{3}{2} atn 
	\; K \right] \right\}
\label{sigma3D}
\end{eqnarray}

In anisotropic thin films we get:

\begin{eqnarray}
	\sigma_{2D}= \frac{e^{2}}{16\hbar 
	d\epsilon}\;\frac{\xi_x(0)}{\xi_y(0)}\; 
	\left(-\frac{2}{w^{2}}\right)\left\{ \left[\frac{1}{2}\;ln 
	\left(1+w^{2}\right)-w\;atn\;w \right]+ \right.  \nonumber \\
	\left.  +i\left[w-\frac{w}{2}\;ln 
	\left(1+w^{2}\right)-atn\left(w\right) \right] 
	-\left[\frac{1}{2}\;ln 
	\left(1+\frac{w^{2}}{(K^{2}+1)^{2}}\right)-w\;atn\left(\frac{w}{K^{2}+1}\right) 
	\right]+ \right.
	\label{sigma2D}
	\\
	\left.  -i\left[\frac{w}{K^{2}+1}-\frac{w}{2}\;ln 
	\left(1+\frac{w^{2}}{(K^{2}+1)^{2}}\right)-atn\left(\frac{w}{K^{2}+1}\right) 
	\right] \right\} \nonumber
\end{eqnarray}

and finally, in thin wires one has:

\begin{equation}
	\sigma_{1D}=\frac{\pi}{16}\;\frac{e^{2}\xi_x(0)}{\hbar 
	L^{2}\epsilon^{3/2}}\;\left(-\frac{16}{\pi w^{2}}\right) \left[ 
	\left(1-\frac{iw}{2}\right)atn\left(K\right) 
	+\frac{iw}{2}\frac{K}{1+K^{2}} 
	-\sqrt{1-iw}\;atn\left(\frac{K}{\sqrt{1-iw}}\right) \right]
\label{sigma1D}
\end{equation}

Eq.s \ref{sigma3D},\ref{sigma2D},\ref{sigma1D} are the main results of 
this paper.\\
A number of already known results can be obtained from these
expressions as limiting cases: taking the limit for
$K\rightarrow\infty$ and isotropic $\xi$'s we recover the Schmidt's
results\cite{schmidt} for the corresponding dimensionalities.  Since
$K=\Lambda/\epsilon^{\frac 1 2}$, the limit $K\rightarrow\infty$
contains either the no-cutoff (usual GL calculation), or the
$T\rightarrow T_{c}$ limits.  Correctly, close to the transition
temperature the usual GL expressions are recovered.\\
As a by-product of Eq.s \ref{sigma3D},\ref{sigma2D},\ref{sigma1D}, it 
is easy to obtain the dc excess conductivity for the various 
dimensionalities in the presence of a short-wavelength cutoff, 
taking the limit $\omega\rightarrow0$.  One gets:

\begin{equation}
	\sigma_{3D}^{dc}=\frac{e^{2}}{32\hbar \xi_z(0)\epsilon^{1/2}} 
	\;\frac{\xi_x(0)}{\xi_y(0)}\; 
	\frac{2}{\pi}\left[atn\;K-K\frac{\frac{5}{3}K^{2}+1}{\left(K^{2}+1\right)^{2}}\right]
\label{dc3D}
\end{equation}

\begin{equation}
	\sigma_{2D}^{dc}=\frac{e^{2}}{16\hbar d\epsilon} 
	\;\frac{\xi_x(0)}{\xi_y(0)}\; \frac{1}{\left(1+K^{-2}\right)^{2}}
\label{dc2D}
\end{equation}

\begin{equation}
	\sigma_{1D}^{dc}=\frac{e^{2}\xi_x(0)}{8\hbar L^{2}\epsilon^{3/2}} 
	\; 
	\left[atn\;K-\frac{K\left(1-K^{2}\right)}{\left(K^{2}+1\right)^{2}}\right]
\label{dc1D}
\end{equation}

Again, a number of already known results are obtained in the 
appropriate limits.  In all cases, for $K\rightarrow\infty$ we get the 
usual results for the (anisotropic) gaussian 
paraconductivity.\cite{skoc} The cutoffed, 3D uniaxially 
isotropic result by Gauzzi\cite{gauzzi} is recovered using 
$\xi_{x}=\xi_{y}$ in Eq.\ref{dc3D} and setting equal cutoffs in 
Ref.\onlinecite{gauzzi}.  Similarly, we recover the known results in 
2D\cite{hopf,gauzzi,silva} and 1D.\cite{coffey}\\

\section{Discussion}
\label{Discussion}

\begin{figure}
\centerline{\psfig{figure=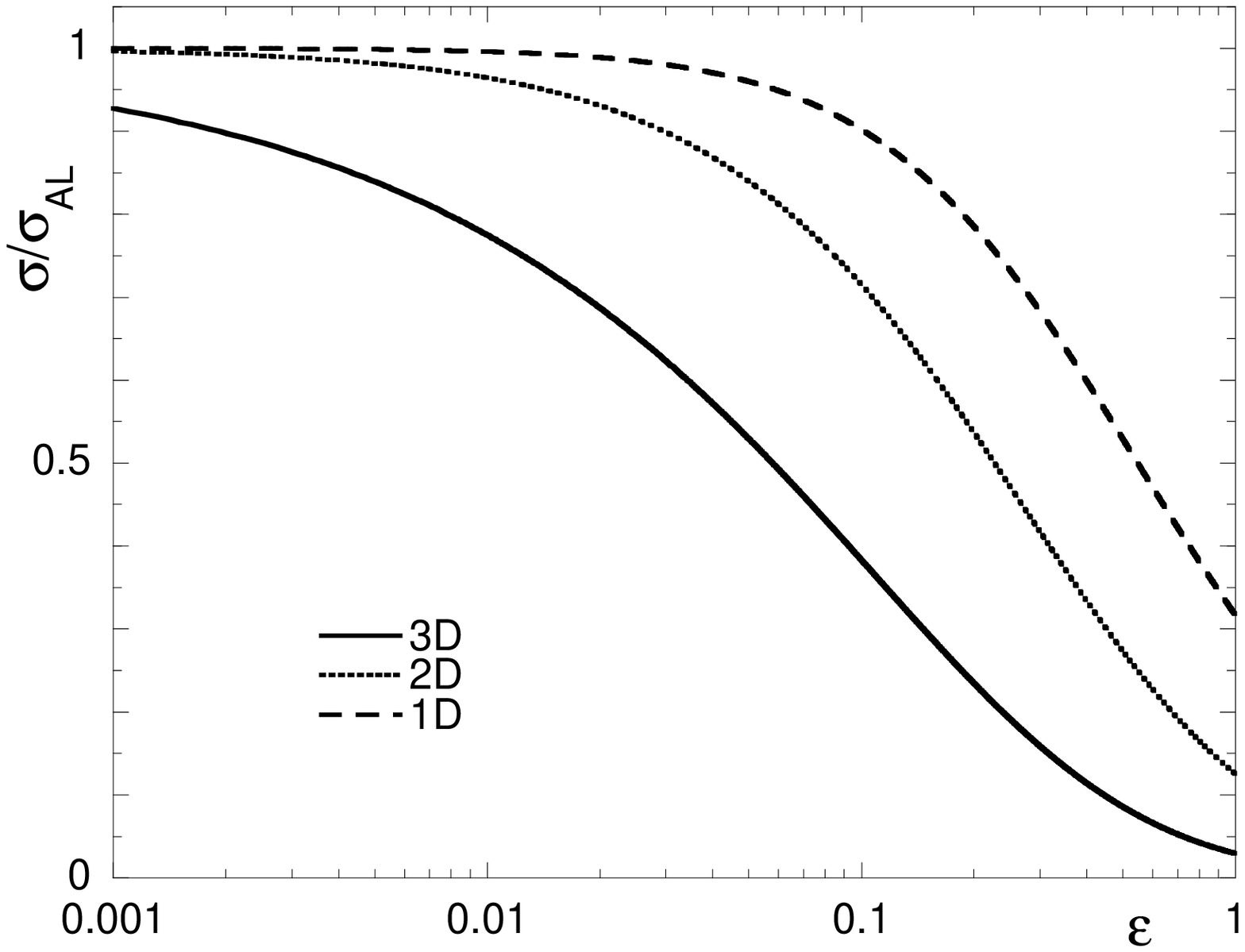,height=9cm,width=12cm,clip=,angle=0.}}

\caption{Correction to the dc conductivity as a function of  the 
reduced temperature $\epsilon$=ln$T/T_{c}$, for cutoff number $\Lambda=0.74$, 
\cite{notaxi} and for different dimensionalities.}
\label{figdc}
\end{figure}

\begin{figure}
\centerline{\psfig{figure=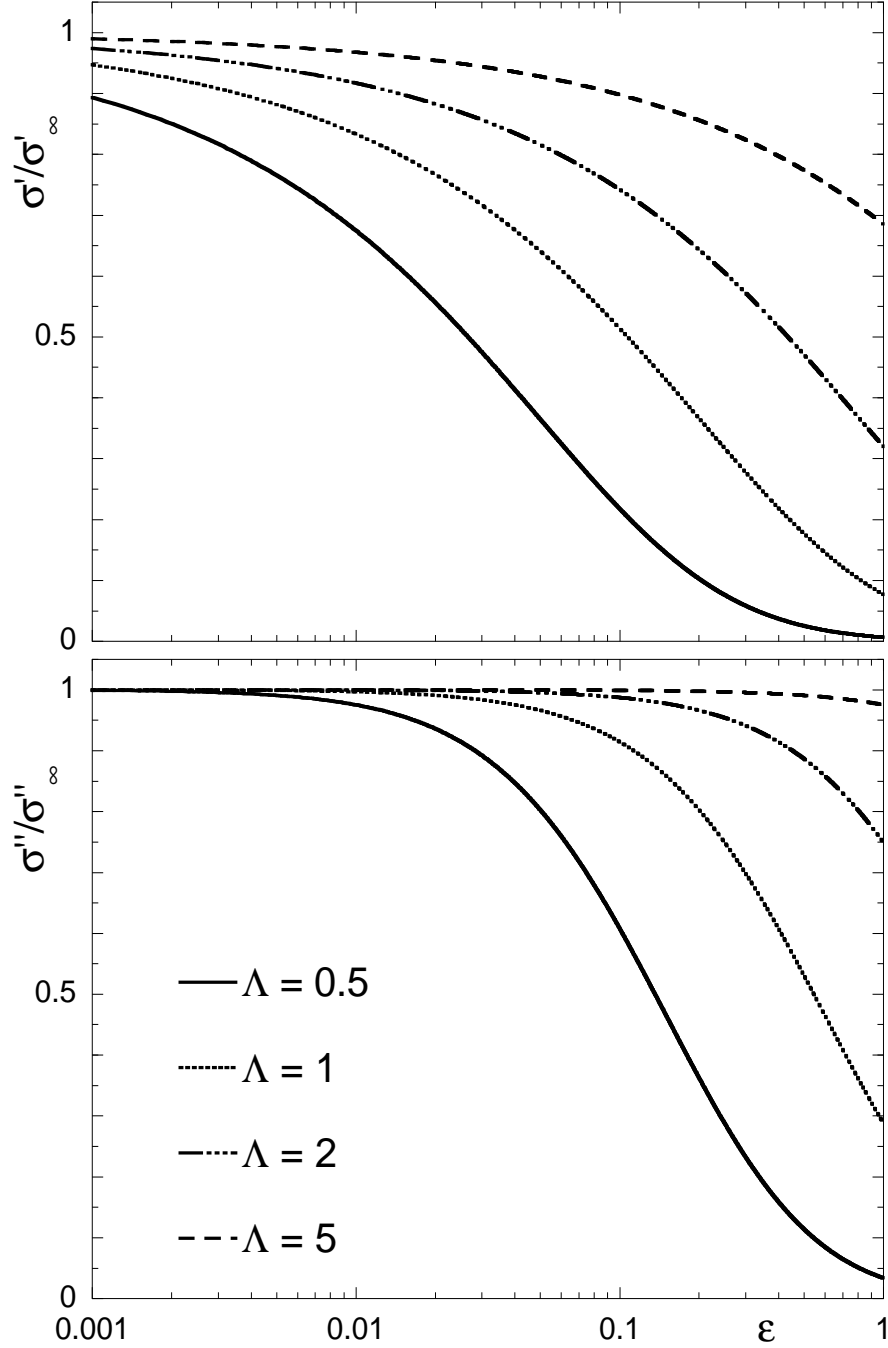,height=18cm,width=12cm,clip=,angle=0.}}

\caption{Correction to the 3D dynamic conductivity vs. the reduced 
temperature $\epsilon$, 
for $\omega\tau_{0}=10^{-3}$ and various cutoff numbers $\Lambda$. Upper panel: 
real part; lower panel: imaginary part. $\tau_{0}/\epsilon$ is the 
Ginzburg-Landau relaxation time (see Section\ref{model}).}
\label{fig3Deps}
\end{figure}

\begin{figure}
\centerline{\psfig{figure=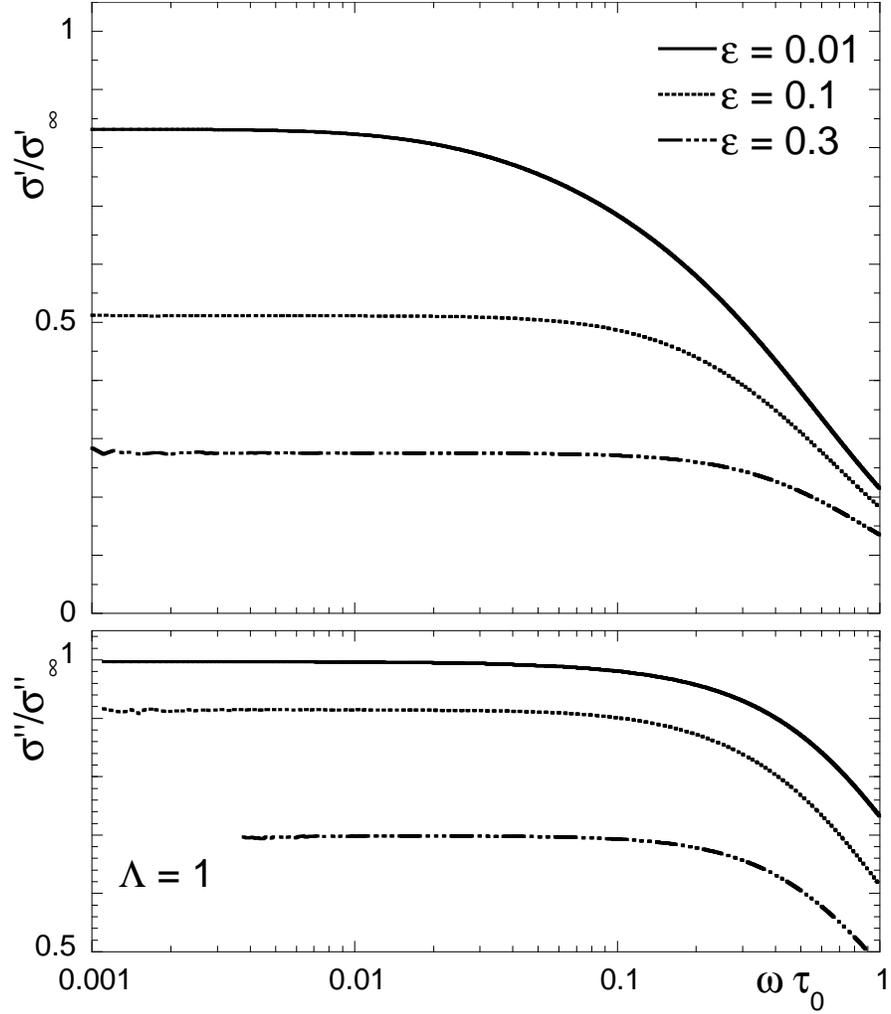,height=13.5cm,width=12cm,clip=,angle=0.}}

\caption{Correction to the 3D dynamic conductivity vs.  
$\omega\tau_{0}$, with cutoff number $\Lambda=1$.  Upper panel: real 
part; lower panel: imaginary part.}
\label{fig3Dot}
\end{figure}

\begin{figure}
\centerline{\psfig{figure=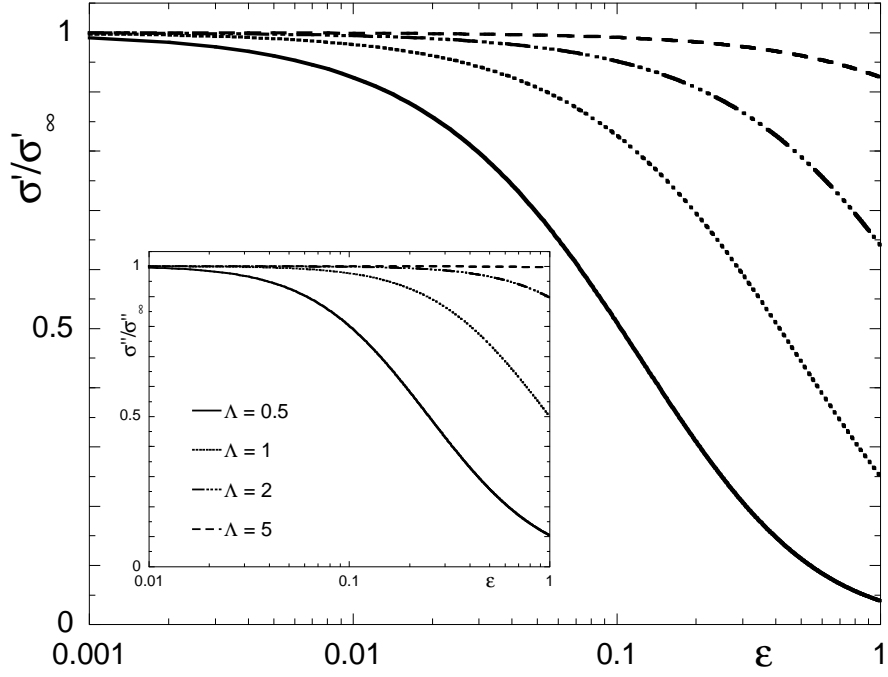,height=9cm,width=12cm,clip=,angle=0.}}

\caption{Correction to the 2D dynamic conductivity vs. the reduced 
temperature $\epsilon$, 
for $\omega\tau_{0}=10^{-3}$ and various cutoff numbers $\Lambda$. Main panel: 
real part; inset: imaginary part.}
\label{fig2Deps}
\end{figure}

\begin{figure}
\centerline{\psfig{figure=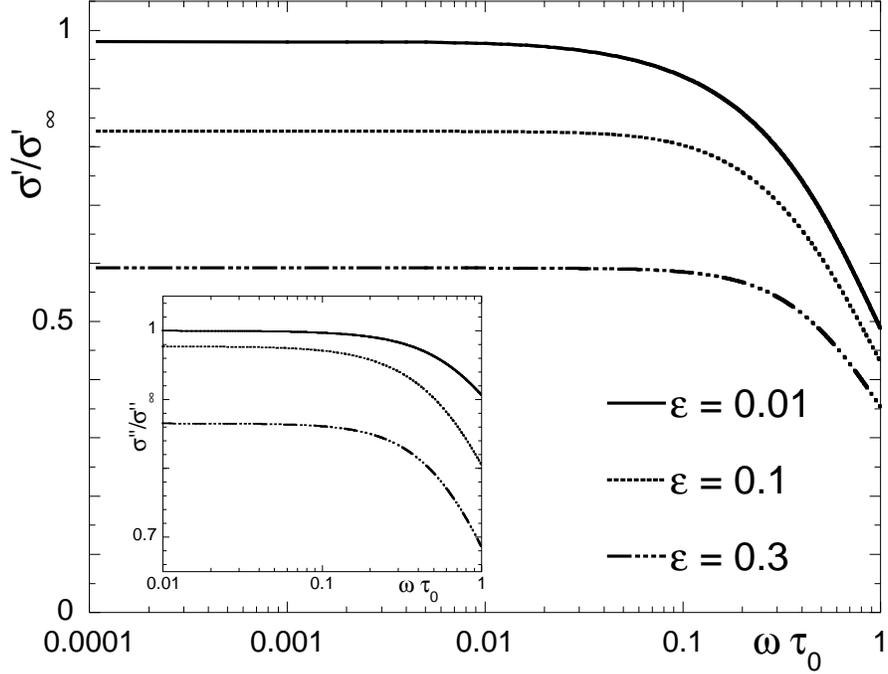,height=9cm,width=12cm,clip=,angle=0.}}

\caption{Correction to the 2D dynamic conductivity vs.  $\omega\tau_{0}$, 
and cutoff number $\Lambda=1$.  Main panel: real part; inset: imaginary part.}
\label{fig2Dot}
\end{figure}

\begin{figure}
\centerline{\psfig{figure=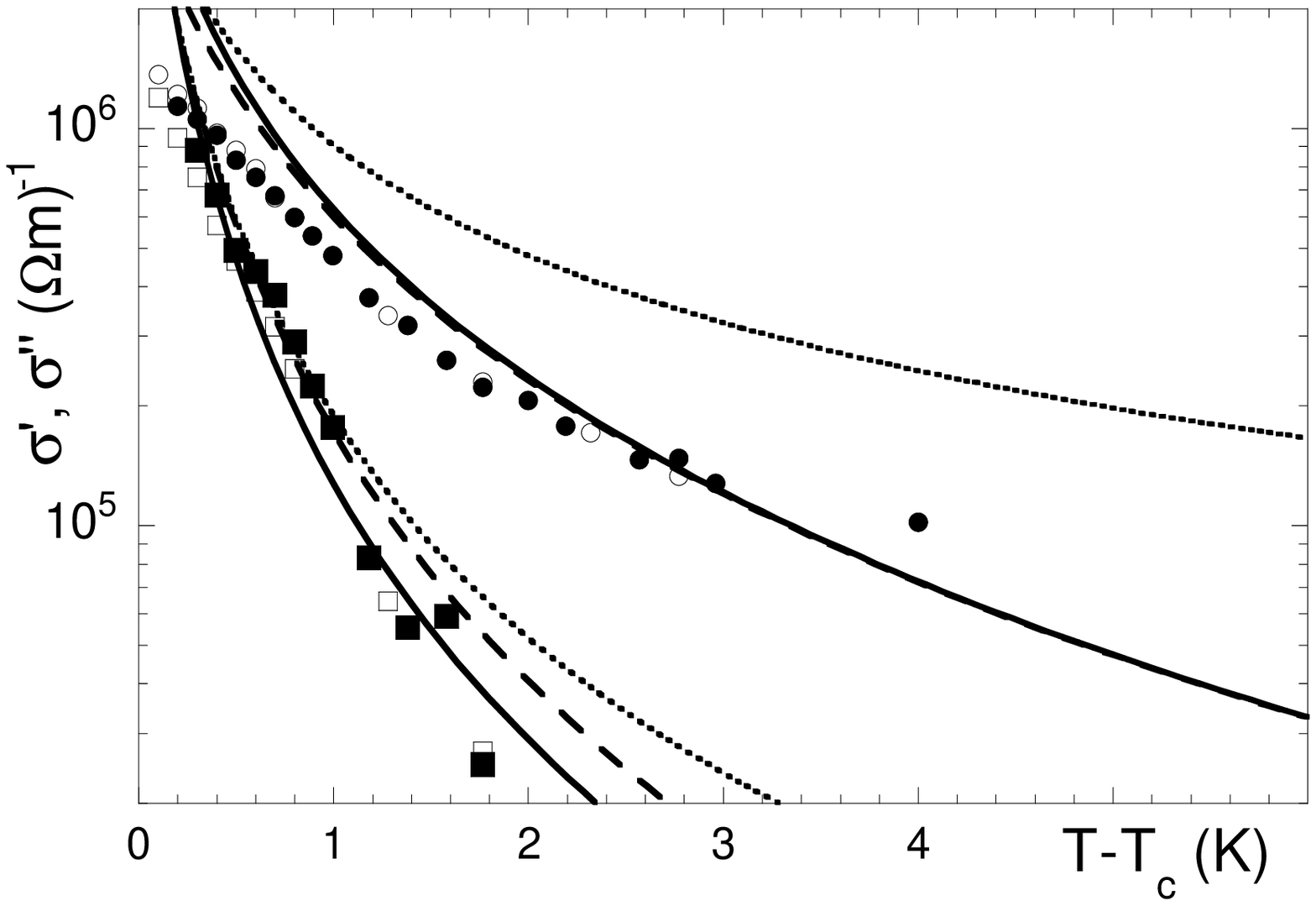,height=9cm,width=12cm,clip=,angle=0.}}

\caption{Data for the dynamic conductivity of Zn-doped YBCO, from 
Ref.\cite{waldram}, at 25 GHz (open symbols) and 36 GHz (full 
symbols).  Circles: real part.  Squares: imaginary part.  Dotted 
lines: fits with the uncutoffed expression, $\omega\tau_{0}=0.0086$, 
corresponding to 36 GHz.  Dashed lines: 2D expression, 
Eq.\ref{sigma2D}, with cutoff number $\Lambda=0.247$ (see text), 
$\omega\tau_{0}=0.0086$.  Continuous lines: 2D expression, 
Eq.\ref{sigma2D}, with $\Lambda=0.247$ (see text), 
$\omega\tau_{0}=0.006$.  The momentum cutoff better describes the 
experimental trends than the plain GL expression.}
\label{figexp}
\end{figure}

There are several interesting features that emerge from the novel 
relations, Eqs.\ref{sigma3D},\ref{sigma2D},\ref{sigma1D} and the 
corresponding dc expressions.  We first discuss shortly some peculiarities of 
the dc conductivity (Eqs.\ref{dc3D},\ref{dc2D},\ref{dc1D}) that seem 
to have been unnoticed in previous works.\\
A plot of the cutoff correction to the gaussian paraconductivity, 
$\sigma^{dc}(\epsilon;\Lambda,\omega=0)/\sigma^{dc}(\epsilon;\infty,\omega=0)$, 
is reported in Fig.\ref{figdc} for $\Lambda$=0.74.\cite{notaxi}  It is apparent 
that the correction is stronger for higher 
dimensionalities.\cite{notadimcros} It is instructive to investigate the cases $K\rightarrow\infty$ (that 
is, the close vicinity of $T_{c}$) and $K<<1$ (that is, high reduced 
temperatures and/or strong cutoff effects).  The latter case is more 
an oversimplification than a physically reachable limit, but it is 
very convenient in order to obtain analytical expressions.  As can be 
derived from Eqs.(\ref{dc3D},\ref{dc2D},\ref{dc1D}), in all the three 
cases one obtains the same temperature dependence, namely:
\begin{equation}
	\sigma^{dc}_{K<<1}=\frac{e^{2}}{\hbar} N_{D} 
	\frac{1}{\epsilon^{3}}
\end{equation}
\label{dcalta}
where $N_{1D}=\frac{1}{3}\frac{\xi_{x}(0)}{L^{2}}\Lambda^{3}$,
$N_{2D}=\frac{1}{32}\frac{\xi_{x}(0)}{d \xi_{y}(0)}\Lambda^{4}$,
$N_{3D}=\frac{1}{30\pi}\frac{\xi_{x}(0)}{\xi_{y}(0)\xi_{z}(0)}\Lambda^{5}$.
 The effect of the cutoff does change with the dimensionality
(different powers of $\Lambda$), but only through a different
prefactor, so that the determination of the dimensionality from the
data at high $\epsilon$ should be carried on with care.  In
particular, it is tempting to identify the dimensionality from some
power law of the excess conductivity with $\epsilon$.  This is surely
not possible far from $T_{c}$, where the power law appears to be the
same for all the dimensionalities.  This fact could be somehow
expected, since the introduction of the SWL cutoff introduces an
additional length in the system, relevant far from the transition. 
Even a quantitative comparison, in the case e.g. of HTCS, might not
allow for unambiguous identification of the dimensionality: taking as
an example $\xi_{x}\simeq\xi_{y}$ and reasonable $\Lambda\sim1$, one
has to compare $d$ with $\pi\xi_{z}(0)$.  In the case of BSCCO, taking
$d\simeq$3.3$\AA$, the thickness of the double CuO layers, and
$\xi_{z}(0)\simeq$1$\AA$, the two prefactors are of the very same
order of magnitude.  In general, only a quantitative comparison in a
wide temperature dependence can give indications on the effective
dimensionality.\\
We now turn to the discussion of the frequency dependence.  We
restrict this part of the discussion to the 2D and 3D cases, since we
are not aware of experimental results for the dynamic conductivity in
1D systems.  We discuss the temperature and frequency dependencies by
means of the reduced variables $\epsilon=ln(T/T_{c})$ and
$\omega\tau_{o}$.  The use of $\omega\tau_{0}$ as a reduced variable
does not require the exact determination of the GL relaxation time. 
However, for numerical estimates, assuming the BCS-derived value for
$\Gamma_{0}$ one has $\omega\tau_{0}\sim10^{-3}$ for $\omega/2\pi$=10
GHz: $\omega\tau_{0}<<1$ even in the microwave region.\\
We employ the useful plots\cite{hopf,coffey} of the conductivity 
normalized to the gaussian (uncutoffed) results (note that by 
using these plots the effects of the anisotropy cancel out).\\
In Fig.\ref{fig3Deps} we plot the real and imaginary part of the
conductivity normalized to the Gaussian values,
$\sigma^{\prime}(\omega,\epsilon,\Lambda)/\sigma^{\prime}(\omega,\epsilon,\infty)$,
and
$\sigma^{\prime\prime}(\omega,\epsilon,\Lambda)/\sigma^{\prime\prime}(\omega,\epsilon,\infty)$,
in the 3D case for various values of the cutoff number, $\Lambda$, as
a function of the reduced temperature $\epsilon$ and
$\omega\tau_{0}=10^{-3}$.  As can be seen, the presence of a SWLC
strongly depresses the dynamic conductivity at high $\epsilon$.  We
note that the correction is much stronger on $\sigma^{\prime}$ than on
$\sigma^{\prime\prime}$.  For cutoff number $\Lambda=1$ and
$\epsilon=10^{-1}$ the real part is reduced by $\sim50\%$ with respect
to the uncutoffed value, while the imaginary part remains at
$\sim90\%$ of the uncutoffed value.  We also note that the effects of
the cutoff are significant even at reduced $\epsilon\sim10^{-2}$, that
is almost in the full accessible temperature range, apart the small
region very close to $T_{c}$ (where, however, additional complications arise
due to the possible entering of the critical region).\\
The frequency dependence is best studied with the plots at fixed 
$\epsilon$ and variable $\omega\tau_{0}$, as reported in 
Fig.\ref{fig3Dot} for three reduced temperatures and $\Lambda=1$.  It 
is easy to notice that for all the $\epsilon$ depicted, the effect of 
the cutoff does not introduce a significant frequency dependence 
neither in $\sigma^{\prime}$, nor in $\sigma^{\prime\prime}$, up to 
very high frequencies (at the high edge of the microwave spectrum).  The same figure 
shows again the larger effect of the cutoff on $\sigma^{\prime}$ than 
in $\sigma^{\prime\prime}$.\\
The fact that, for reasonable cutoff numbers and frequencies, the 
frequency dependence of the correction is nearly flat, while the 
temperature dependence changes with respect to the uncutoffed value, 
has a rather intriguing consequence: in fact, as recalled in 
Sec.\ref{model}, the gaussian dynamic conductivity can be cast in a 
scaling form: $\sigma(\omega,\epsilon)=\sigma_{dc}(\epsilon) 
S(\omega\tau(\epsilon))$, where $\tau(\epsilon)=\tau_{0}/\epsilon$ in 
the gaussian regime.  Our present results still predict an {\em 
approximate} scaling, but the temperature dependence of $\tau$ is 
apparently different from the uncutoffed result.\cite{notaspiega} The 
exact scaling prediction\cite{dorsey} is reinstated for infinite 
cutoff. It 
would be interesting to perform wideband measurements in an extended 
temperature range above $T_{c}$ in view of this novel feature of the 
theoretical expressions.\\
Essentially the same features, but with a less pronounced effect of 
the cutoff are exhibited by 2D systems, as shown by Fig.s \ref{fig2Deps} and 
\ref{fig2Dot}. We then expect that in strongly layered systems the 
frequency dependence of the
dynamic paraconductivity is essentially unaffected by the cutoff in a 
resonably wide temperature range (up to $\epsilon\sim0.2$), while effects 
of the cutoff should be easier to observe in measurements of the real part of the 
paraconductivity at fixed frequencies and variable $\epsilon$.\\
We now discuss our results in comparison with experimental data for 
the microwave excess conductivity in high-$T_{c}$ 
superconductors.  It should be noted that most of the microwave data, 
where one might observe some significant effect of the frequency, were 
devoted to studies well inside the superconducting state or, at most, 
of the critical regime very close to $T_{c}$.  As a consequence, there 
are nearly no published data (to our knowledge) of the 
finite-frequency excess conductivity, in temperature ranges extended 
enough above $T_{c}$.  In order to assess the question wether or not 
our cutoff approach describes the correct trend in the observations, 
we consider the data for the complex excess conductivity above $T_{c}$ 
taken in Zn-doped YBCO crystals.\cite{waldram} In this case, a 
Ginzburg-Landau-like analysis suggested that the compound behaves 
nearly as a 2D system,\cite{waldram} but the overall magnitude of the 
excess conductivity was about half of the theoretical prediction when 
the cell unit ($\approx$11.7\AA) was taken as effective layer 
separation.  In Fig.\ref{figexp} we plot the measured complex excess 
conductivity at 25 and 36 GHz (as digitized from Ref.\onlinecite{waldram}), 
the 2D GL prediction ($\Lambda=\infty$) at 36 GHz and the 2D, cutoffed 
curve at the two frequencies for $\Lambda=0.247$ (that 
is,\cite{notaxi} a momentum cutoff of $(3\xi_{0})^{-1}$).  The result is 
encouraging: in particular, a nearly quantitative fit is obtained not 
too close to $T_{c}$ (we do not discuss the possible occurence of a 
dimensional crossover approaching $T_{c}$), without adjusting the other 
parameters,\cite{scalefactor} thus demonstrating that the proposed 
extension of the GL theory can correctly describe the features 
experimentally observed.  New measurements, in extended temperature 
ranges above $T_{c}$, would be extremely useful to check the 
applicability of the momentum-cutoff scenario.\\

\section{Conclusions}
\label{conc}
In conclusion, we have performed a Ginzburg-Landau calculation of the 
dynamic conductivity in anisotropic bulk, thin films and thin wires 
superconductors.  According to the general prescriptions of the GL 
theory,\cite{mishonov,werthamer} we have introduced a short-wavelength 
cutoff in the spectrum of the accessible momenta.  We have derived 
closed forms for the dynamic conductivity in the three cases 
abovementioned, that can be directly used for data fitting.  We have 
predicted peculiar behaviors of the complex conductivity as the 
short-wavelength regime shows up.  The analysis of some of the 
existing data for the finite-frequency conductivity finds a 
description in terms of the model here presented. More data are 
required to discuss the details of the excess conductivity at 
microwave frequencies.

\section*{acknowledgements}
We thank R.Raimondi and S.Sarti for useful discussions, D.Neri for 
stimulating our interest in this field and for many useful discussions in the 
early stage of this work, M.W.Coffey for sending Ref.\onlinecite{coffey} 
prior to publication and for useful discussions, and T.Mishonov for 
stimulating correspondence and useful suggestions.  \\

\appendix
\section*{Additional remarks: the total-energy cutoff}
It is a recently debated topic wether the momentum cutoff here adopted 
should be or not replaced by a cutoff in the total energy of the 
pair.\cite{mishonov,carballeira} In the 3D energy cutoff scenario, the 
number $\Lambda$ should be replaced by $\sqrt{c^{2}-\epsilon}$, with $c$ a 
new cutoff number, so that momentum-cutoff expressions can be 
easily rewritten in terms of energy cutoff (for a discussion, we 
address the reader to the review, Ref.\onlinecite{mishonov}).  Measurements 
of the dc conductivity showed a quite good agreement with the momentum 
cutoff scenario in BSCCO crystals,\cite{silvaPhC,silva} and with the 
energy cutoff in YBCO crystals.\cite{carballeira} Analysis of the 
finite-frequency conductivity at high reduced temperatures might help 
to resolve this issue.

\newpage

\newpage

\newpage

\newpage

\newpage

\begin{references}

\bibitem{mishonov} T.Mishonov and E.Penev, {\em Int.  J. Mod.  Phys B} {\bf 
14}, 3831 (2001).

\bibitem{skoc} W.J.Skocpol and M.Tinkham, {\em Rep.  Prog.  Phys.}
{\bf 38}, 1049 (1975).

\bibitem{freitas} P.P.Freitas, C.C.Tsuei, and T.S.Plaskett, {\em Phys.  
Rev.  B} {\bf 36}, 833 (1987).

\bibitem{ausloos} M.Ausloos and Ch.Laurent, {\em Phys.  Rev.  B}
{\bf 37}, 611 (1988).

\bibitem{hopf} R.Hopfengartner, B.Hensel and G.Saemann-Ischenko, {\em 
Phys.  Rev.  B} {\bf 44}, 741 (1991).

\bibitem{balest92} G.Balestrino, M.Marinelli, E.Milani, {\em Phys.  
Rev.  B} {\bf 46}, 14919 (1992).

\bibitem{labdi} S.Labdi, S.Megtert, and H.Raffy, {\em Solid State 
Commun.} {\bf 85}, 491 (1993).

\bibitem{calzona} V.Calzona, M.R.Cimberle, C.Ferdeghini, G.Grasso, 
D.V.Livanov, D.Marr\'{e}, M.Putti, A.S.Siri, G.Balestrino, and 
E.Milani, {\em Solid State Commun.} {\bf 87}, 397 (1993).

\bibitem{pavuna} A.Gauzzi and D.Pavuna {\em Phys.  Rev.  B} {\bf 51}, 
15420 (1995).

\bibitem{asla} L.G.Aslamazov and A.I.Larkin, {\em Soviet Phys.  Solid 
State} {\bf 10}, 875 (1968) [{\em Fiz.  Tverd.  Tela} (Leningrad) {\bf 
10}, 1104 (1968)]; {\em Phys.  Lett.} {\bf 26A}, 238 (1968).

\bibitem{schmidt} H.Schmidt, {\em Z. Phys } {\bf 216}, 336 (1968); 
ibidem {\bf 232}, 443 (1970).

\bibitem{maki} K. Maki, {\em Prog.  Theor.  Phys.} {\bf 39}, 897 (1968); 
ibidem {\bf 40}, 193 (1968).

\bibitem{klemm} R.A.Klemm, {\em J. Low Temp.  Phys.} {\bf 16}, 381 
(1974).

\bibitem{reggiani} L.Reggiani, R.Vaglio, A.A.Varlamov, {\em Phys.  
Rev.  B} {\bf 44}, 9541 (1991).

\bibitem{gauzzi} A.Gauzzi, {\em Europhys.  Lett.}
{\bf 21}, 207 (1993).

\bibitem{silvaPhC} E.Silva, D.Neri, M.Esposito, R.Fastampa, M.Giura, S.Sarti, {\em 
Physica C} {\bf 341-348}, 1927 (2000).

\bibitem{silva} E.Silva, S.Sarti, R.Fastampa, M.Giura, {\em Phys.  
Rev.  B} {\bf 64}, 144508 (2001).

\bibitem{thom} R.S.Thompson, {\em Phys.  Rev.  B} {\bf 1}, 
327 (1970).

\bibitem{johnson2} W.L.Johnson and C.C.Tsuei, {\em 
Phys.  Rev.  B} {\bf 13}, 4827  (1976).

\bibitem{johnson} W.L.Johnson, C.C.Tsuei, and P.Chaudhari, {\em Phys.  
Rev.  B} {\bf 17}, 2884 (1978).

\bibitem{werthamer} N.R.Werthamer, in {\em Superconductivity}, ed.  by 
R.D.Parks (Marcel Dekker, New York, 1969), p.321-370.

\bibitem{patton} B.R.Patton, V.Ambegaokar, and J.Wilkins, {\em Solid 
State Comm.} {\bf 7}, 1287 (1969).

\bibitem{gollub2} J.P.Gollub, M.R.Beasley, and M.Tinkham, 
{\em Phys.  Rev.  Lett.} {\bf 25}, 1646 (1970).

\bibitem{gollub} J.P.Gollub, M.R.Beasley, R.Callarotti and M.Tinkham, 
{\em Phys.  Rev.  B} {\bf 7}, 3039 (1973).

\bibitem{magnetic} see, e.g., M.E.Fisher and J.S.Langer, 
{\em Phys.  Rev.  Lett.} {\bf 20}, 665 (1968); D.J.W.Geldart, 
T.G.Richard, {\em Phys.  Rev.  B} {\bf 12}, 5175 (1975); M.Ausloos and 
K.Durczewski, {\em Phys.  Rev.  B} {\bf 22}, 2439 (1980); and 
references therein.

\bibitem{nerisatt} D.Neri, R.Marcon, E.Silva, R.Fastampa, L.Iacobucci, 
S.Sarti, {\em Int.  J. of Mod.  Phys.  B} {\bf 13}, 1097 (1999)

\bibitem{nerimos} D.Neri, R.Fastampa, M.Giura, R.Marcon, S.Sarti, 
E.Silva, {\em J. of Low Temp.  Phys.} {\bf 117}, 1099 (1999)

\bibitem{coffey} M.W.Coffey, preprint (2001).

\bibitem{makitaka} K.Maki and H.Takayama, {\em J. Low Temp. Phys.} {\bf 
5}, 313  (1971).

\bibitem{kurk} J.Kurkijarvi, V.Ambegaokar, and G.Eilemberger, {\em 
Phys.  Rev.  B} {\bf 5}, 868 (1972).

\bibitem{gerhardts} R.R.Gerhardts, {\em Phys.  Rev. B}
{\bf 9}, 2945 (1974).

\bibitem{varlareview} A.A.Varlamov, G.Balestrino, E.Milani and 
D.V.Livanov, {\em Adv.  in Phys.} {\bf 48}, 655 (1999).

\bibitem{carballeira} C.Carballeira, S.R.Curr\'{a}s, J.Vina, 
J.A.Veira, M.V.Ramallo, F.Vidal, {\em Phys.  
Rev.  B} {\bf 63}, 144515 (2001).

\bibitem{wang} Q.Wang, G.A.Saunders, H.J.Liu, M.S.Acres, and 
D.P.Almond, {\em Phys.  Rev.  B} {\bf 55}, 8529 (1997).

\bibitem{oldmicro} R.V.D'Aiello and S.J.Freedman, {\em Phys. Rev. Lett.} 
{\bf 22}, 515 (1969); S.L.Lehoczky and C.V.Briscoe {\em Phys.  Rev.  
Lett.} {\bf 23}, 695 (1969); {\em Phys.  Rev.  B} {\bf 4}, 3938 
(1971).
\bibitem{anlage} S.M.Anlage, J.Mao, J.C.Booth, D.H.Wu and J.L.Peng 
{\em Phys.  Rev.  B} {\bf 53}, 2792 (1996).

\bibitem{booth} J.C.Booth, D.H.Wu, S.B.Qadri, E.F.Skelton, 
M.S.Osofsky, A.Piqu\'{e} and S.M.Anlage, {\em Phys.  Rev.  Lett.} 
{\bf 77}, 4438 (1996).

\bibitem{neri} D.Neri, E.Silva, S.Sarti, R.Marcon, M.Giura, 
R.Fastampa, N.Sparvieri, {\em Phys.  Rev.  B} {\bf 58}, 14581 (1998).
 
\bibitem{waldram} J.R. Waldram, D.M. Broun, D.C. Morgan, R. Ormeno, A. 
Porch, {\em Phys.  Rev.  B} {\bf 59}, 1528 (1999).

\bibitem{dorsey} A.T.Dorsey, {\em Phys.  Rev.  B} {\bf 43}, 7575 
(1991).

\bibitem{ausMT} M.Ausloos, F.Gillet, Ch.  Laurent, P.Clippe {\em Z. 
Phys.  B} {\bf 84}, 13 (1991).

\bibitem{livanov} D.V.Livanov, G.Savona, and A.A.Varlamov, {\em Phys.  
Rev.  B} {\bf 62}, 8675 (2000).

\bibitem{gorkov} L.P.Gor'kov, {\em Zh.  Eksp.  Teor.  Fiz.} {\bf 36}, 
1918 (1959) [{\em Soviet Phys.  JETP} {\bf 9}, 1364 (1959)].

\bibitem{mikeska} H.-J.Mikeska and H.Schmidt, {\em Z. Physik} {\bf 
230}, 239 (1970).

\bibitem{gamma0} In BCS superconductors one finds,\cite{skoc} from 
comparison with microscopic theory, $\Gamma_0=\left(8k_BT / \hbar \pi 
a \right)$, where $k_{B}$ is the Boltzmann constant. In 
exotic superconductors the value of $\Gamma_0$ may be numerically 
different.  This aspect has been clearly discussed in 
Ref.\onlinecite{mishonov}. The qualitative results here presented do not depend on 
the precise numerical value of $\Gamma_{0}$.

\bibitem{notamisho} The convergence of the integral written in this form 
is assured when expansion for small fields is made in the subsequent 
steps of the calculation. The full expression is obviously convergent, 
see Eq.(4) in Ref.\onlinecite{neri}. We thank T. Mishonov for bringing our 
attention to this point.

\bibitem{notaxi} For the sake of compactness, throughout the paper we 
use only the GL zero-temperature correlation length, $\xi(0)$, instead 
of the microscopic coherence length $\xi_{0}=\xi(0)/0.74$.  As a 
consequence, the cutoff $\Lambda$=0.74 means a momentum cutoff 
$Q=\xi_{0}^{-1}$

\bibitem{ld} W.E.Lawrence and S.Doniach, {\em Proceedings of the 12th 
International Conference on Low Temperature Physics}, edited by 
E.Kanda (Kiegaku, Tokio, 1971), pp.361-362.
 
\bibitem{notadimcros} As pointed out by other authors,\cite{hopf,pavuna,gauzzi,coffey} 
clear changes of slopes are observed, that can easily mimic the 
appearance of dimensional crossovers. 

\bibitem{notaspiega} In fact, it is easy to derive an approximate 
expression for $\sigma$ with $w<<1$, from which both the extremely 
slow frequency dependence and the new approximate scaling are exploited.

\bibitem{scalefactor}  A scale factor would substantially 
improve the fit, at the expense of the introduction of an additional 
free parameter. Purpose of the present comparison with the data is to 
show that the proposed extension to the GL theory is in agreement with 
the experimental trends. 

\end{references}
\end{document}